\documentclass[prl,twocolumn,showpacs,a4paper]{revtex4}
\usepackage{amssymb}
\usepackage{epsfig}
\usepackage{amsmath}
\usepackage{graphicx}
\usepackage{bm}

\begin{document}

\title {Emergence of atom-light-mirror entanglement inside an optical cavity}
\author{C. Genes, D. Vitali and P. Tombesi}
\affiliation{Dipartimento di Fisica, Universit\`{a} di Camerino, I-62032 Camerino (MC),
Italy}

\begin{abstract}
We propose a scheme for the realization of a hybrid, strongly quantum-correlated system formed of an atomic ensemble surrounded by a
high-finesse optical cavity with a vibrating mirror. We show that the steady state of the system shows tripartite and bipartite continuous
variable entanglement in experimentally accessible parameter regimes, which is robust against temperature.
\end{abstract}

 \maketitle

Recently there has been an increasing convergence between condensed matter physics and quantum optics, which has manifested in different ways.
On one hand, systems of cold trapped atoms \cite{bloch}, ions \cite{cirac} and electrons \cite{ciara} may realize quantum simulators able to
reproduce and study condensed matter concepts such as Fermi surfaces and Heisenberg models in a controllable and tunable way. On the other hand,
circuit cavity QED \cite{wallraff} provides an example where nano- and micro-structured condensed matter systems are specifically designed in
order to reproduce the phenomena and control of quantum coherence typical of quantum optics system. Alternatively, one can design schemes in
which one has a direct, strong coupling between an atomic degree of freedom and a condensed matter system. Examples of this latter kind are
ion-nanomechanical oscillator \cite{tian1}, or ion-Cooper-pair box \cite{tian2} systems, or a Bose-Einstein condensate coupled to a cantilever
via a magnetic tip \cite{Reichel07}. A further important example is provided by cavity optomechanical systems for which strong coupling between
an optical cavity mode and a vibrational mode by radiation pressure has been already demonstrated
\cite{cohadon99,karrai04,vahala1,gigan06,arcizet06,arcizet06b,bouwm,vahalacool,mavalvala,rugar,harris}, and for which schemes able to show
quantum entanglement \cite{Manc02,pinard05,pir06,prl07} and even quantum teleportation \cite{prltelep} have been already proposed. In these
systems, the radiation pressure interaction can be made considerably large so that genuine quantum effects can be realized when microcavities
and extremely light acoustic resonators \cite{vahala1,gigan06,arcizet06,arcizet06b,bouwm,vahalacool,harris} are used.

\begin{figure}[b]
\centerline{\includegraphics[width=0.45\textwidth]{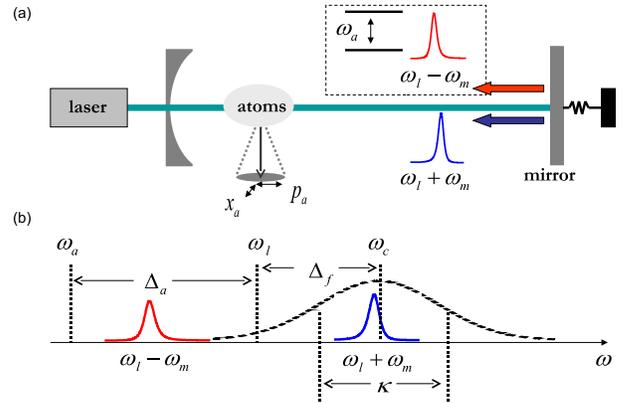}} \caption{(Color online) (a) The cavity is driven by a laser at frequency
$\omega_l$ and the moving mirror at frequency $\omega_m$ scatters photons on the two sidebands at frequency $\omega_l \pm \omega_m$. (b) If the
cavity with frequency $\omega_c$ and bandwidth $\kappa$, is put into resonance with the Antistokes sideband (blue), outgoing cavity photons cool
the mirror vibrational mode. If the atoms are off-resonance with the cavity but resonantly coupled to the red sideband, an entangled tripartite
atom-field-mirror system emerges.} \label{scheme}
\end{figure}

In this letter, we propose a hybrid system formed by an atomic ensemble placed within an optical Fabry-Perot cavity, in which a micromechanical
resonator represents one of the mirrors [see Fig. \ref{scheme}(a)]. The atoms are indirectly coupled to the mechanical oscillator via the common
interaction with the intracavity field. As a first step towards quantum state engineering of mechanical oscillators and quantum state transfer
between atoms and mirrors, we show that using state-of-the-art technology it is possible to generate stationary and robust continuous variable
(CV) tripartite entanglement in the field-atoms-mirror system. To this purpose, we consider $N_{a}$ two-level atoms placed in an optical cavity
under weak-coupling conditions and far from the cavity main resonance $\omega _{c}$. CV tripartite entanglement can be generated by choosing as
working point for the optical cavity with vibrating mirror, the parameter regime corresponding to the ground state cooling of the mechanical
resonator \cite{kippenberg07,girvin07,genes07,dantan07}. In fact, preferential scattering of cavity light into a higher frequency motional
sideband of the driving laser is responsible for cooling of the mechanical system. It has been shown in \cite{prl07} that in this cooling
regime, field-mirror entanglement can be generated, which can be explained in terms of sideband scattering because such an entanglement is
mostly carried by the Stokes sideband. All these facts are at the basis of the robust tripartite atom-resonator-field entanglement reported
here. In fact, if the laser anti-Stokes sideband is resonant with the cavity, the mechanical resonator is cooled by photon leakage and if then
the atomic frequency matches the red (Stokes) sideband frequency, a resonant atoms-mirror coupling mediated by the cavity field is established.
We shall see that in such a regime robust CV tripartite and bipartite entanglement is generated.

\textit{Description of the system}. We consider an optical cavity with a fixed input mirror and a second oscillating mirror, which is driven by
a laser at frequency $\omega_l$. An ensemble of two-level atoms is placed inside the cavity and it is off-resonantly coupled by a collective
Tavis-Cummings type interaction to the optical field \cite{Tavis}. Mirror vibrational motion can be modeled by a harmonic oscillator of
frequency $\omega _{m}$ and decay rate $\gamma _{m}$. In the absence of dissipation and fluctuations the total Hamiltonian of the system is
given by the sum of a free evolution term
\begin{equation}\label{ham0}
H_{0} =\hbar \omega _{c}a^{\dag }a+\frac{\hbar \omega _{a}}{2}S_{z}+\frac{\hbar \omega _{m}}{2}(q^{2}+p^{2}),
\end{equation}
and the interaction term
\begin{eqnarray}H_{I} &=&\hbar g\left( S_{+}a+S_{-}a^{\dag }\right) -\hbar G_{0}a^{\dag }aq \nonumber \\
&&+i\hbar E_{l}\left( a^{\dag }e^{-i\omega _{l}t}-ae^{i\omega _{l}t}\right) . \label{ham1}
\end{eqnarray}
The laser drives significantly only a single cavity mode with frequency $\omega_c$, bandwidth $\kappa$ and annihilation operator $a$ (with
$\left[ a,a^{\dag }\right] =1$). The atomic ensemble is comprised of $N_{a}$
two-level atoms with natural frequency $\omega _{a}$ each described by the $%
1/2$ spin algebra of Pauli matrices $\sigma _{+},\sigma _{-}~$\ and $\sigma
_{z}$. Collective spin operators are defined as $S_{+,-,z}=\sum_{\{i\}}%
\sigma _{+,-,z}^{(i)}$ for $i=1,N_{a}$ and satisfy the commutation
relations $\left[ S_{+},S_{-}\right] =S_{z}$ and $\left[ S_{z},S_{\pm }%
\right] =\pm 2 S_{\pm }$. The mechanical mode dimensionless position and momentum operators $q$ and $p$ satisfy $\left[ q,p\right] =i$. The
atom-cavity coupling constant is given by $g=\mu \sqrt{\omega_c/2\hbar \epsilon_0 V}$ where $V$ is the cavity mode volume and $\mu$ is the
dipole moment of the atomic transition. The radiation pressure coupling constant is instead given by $G_0=(\omega_c/L)\sqrt{\hbar/m \omega_m}$,
where $m$ is the effective mass of the mechanical mode, and $L$ is the length of the cavity. The last term describes the driving of the cavity
by the laser with amplitude $E_{l}$, which is related to the input power $P_{l}$ and the cavity decay rate $\kappa $ by $\left\vert
E_{l}\right\vert =\sqrt{2P_{l}\kappa /\hbar \omega _{l}}$.

The dynamics of the tripartite atom-field-mirror system can be described by a set of nonlinear Langevin equations in which dissipation and
fluctuation terms are added to the Heisenberg equations of motion derived from the Hamiltonian of Eqs.~(\ref{ham0})-(\ref{ham1}) \cite{gard}.
However, we consider a simplified version of such equations, which is valid in the low atomic excitation limit, i.e., when all the atoms are
initially prepared in their ground state, so that $S_z \simeq \left\langle S_{z}\right\rangle \simeq -N_{a}$ and this condition is not
appreciably altered by the interaction with the cavity. This is satisfied when the excitation probability of a single atom is small. In this
limit the dynamics of the atomic polarization can be described in terms of bosonic operators: in fact if one defines the atomic annihilation
operator $c=S_{-}/\sqrt{\left\vert \left\langle S_{z}\right\rangle \right\vert }$, one can see that it satisfies the usual bosonic commutation
relation $[c,c^{\dag }]=1$ \cite{holstein40}. In the frame rotating at the laser frequency $\omega_l$ for the atom-cavity system, the quantum
Langevin equations can then be written as
\begin{subequations}
\label{QLEs}
\begin{align}
\dot{q}& =\omega _{m}p, \\
\dot{p}& =-\omega _{m}q-\gamma _{m}p+G_{0}a^{\dag }a+\xi , \\
\overset{\cdot }{a}& =-(\kappa +i\Delta _{f})a+iG_{0}aq-iG_{a}c+E_{l}+\sqrt{%
2\kappa }a_{in}, \\
\overset{\cdot }{c}& =-\left(\gamma _{a}+i\Delta _{a}\right)c-iG_{a}a+\sqrt{2\gamma_a}F_{c},
\end{align}%
\end{subequations}
where $\Delta _{f}=\omega _{c}-\omega _{l}$ and $\Delta _{a}=\omega _{a}-\omega _{l}$ are respectively the cavity and atomic detuning with
respect to the laser, $G_a=g\sqrt{N_a}$, and $2\gamma_a$ is the decay rate of the atomic excited level. The Langevin noise operators affecting
the system have zero mean value, the Hermitian Brownian noise operator $\xi$ has correlation function $ \langle \xi(t) \xi(t') \rangle =
(\gamma_m/2\pi\omega_m) \int
  d\omega e^{-i\omega(t-t')} \omega [\coth (\hbar \omega/2k_BT)+1]
$ ($k_B$ is the Boltzmann constant and $T$ the temperature of the mechanical oscillator reservoir) \cite{GIOV01}, while the only nonvanishing
correlation function of the noises affecting atoms and cavity field is $\langle a_{in}\left( t\right) a_{in}^{\dagger }\left( t^{\prime }\right)
\rangle =\langle F_{c}\left( t\right) F_{c}^{\dagger }\left( t^{\prime }\right) \rangle=\delta \left( t-t^{\prime }\right) $.

We now assume that the cavity is intensely driven, so that at the steady state, the intracavity field has a large amplitude $\alpha_s$, with
$|\alpha_s| \gg 1$. However, the single-atom excitation probability is $g^2|\alpha_s|^2/(\Delta_a^2+\gamma_a^2)$ and since this probability has
to be much smaller than one for the validity of the bosonic description of the atomic polarization, this imposes an upper bound to $|\alpha_s|$.
Therefore the two conditions are simultaneously satisfied only if \emph{the atoms are weakly coupled to the cavity}, $g^2 \ll \Delta_a^2
+\gamma_a^2$.

In the strong-driving limit, one has a semiclassical steady state; the corresponding mean values can be determined by setting the time
derivatives to zero and factorizing the averages in Eqs.~(\ref{QLEs}), and then solving the corresponding set of nonlinear algebraic equations.
The resulting stationary values are $p_{s}=0$, $q_{s}=G_{0}\left\vert \alpha _{s}\right\vert ^{2}/\omega _{m}$, $c_{s}=-iG_a\alpha _{s}/\left(
\gamma _{a}+i\Delta _{a}\right) $, where the stationary intracavity field is the solution of the nonlinear equation
$\alpha _{s}\left[ \kappa +i\Delta_f-iG_0^2 |\alpha_s|^2/\omega_m +G_{a}^{2}/(\gamma_a +i\Delta _{a}) \right] =E_l $.
We are interested in establishing the presence of quantum correlations among atoms, field and mirror, at the steady state. This can be done by
analyzing the dynamics of the quantum fluctuations of the system around the steady state. It is convenient to consider the vector of quadrature
fluctuations $u=\left( \delta q,\delta p,\delta X,\delta Y,\delta x,\delta y\right) ^{\intercal }$, where $\delta X\equiv(\delta a+\delta
a^{\dag})/\sqrt{2}$, $\delta Y\equiv(\delta a-\delta a^{\dag})/i\sqrt{2}$, $\delta x\equiv(\delta c+\delta c^{\dag})/\sqrt{2}$, and $\delta
y\equiv(\delta c-\delta c^{\dag})/i\sqrt{2}$, and linearize the quantum Langevin equations (\ref{QLEs}) around the steady state values. The
resulting evolution equation for the fluctuation vector is
\begin{equation}
\dot{u}=Au+n,
\end{equation}%
where the drift matrix $A$ is given by
\begin{equation}
A=%
\begin{pmatrix}
0 & \omega _{m} & 0 & 0 & 0 & 0 \\
-\omega _{m} & -\gamma _{m} & G_{m} & 0 & 0 & 0 \\
0 & 0 & -\kappa & \Delta & 0 & G_{a} \\
G_{m} & 0 & -\Delta & -\kappa & -G_{a} & 0 \\
0 & 0 & 0 & G_{a} & -\gamma _{a} & \Delta _{a} \\
0 & 0 & -G_{a} & 0 & -\Delta _{a} & -\gamma _{a}%
\end{pmatrix}%
\end{equation}%
with the effective optomechanical coupling $G_{m}=G_{0}\alpha _{s}\sqrt{2}$ (we have chosen the phase reference so that $\alpha_s$ can be taken real)
and the effective cavity detuning $\Delta=\Delta_f-G_m^2/2\omega_m$. The vector of noises $n$ is given by $n=\left( 0,\xi ,\sqrt{2\kappa }X_{in},\sqrt{%
2\kappa }Y_{in},\sqrt{2\gamma _{a}}x_{in},\sqrt{2\gamma _{a}}y_{in}\right) ^{\intercal }$, where $X_{in}=( a_{in}+a_{in}^{\dag }) /\sqrt{2},$
$Y_{in}=( a_{in}-a_{in}^{\dag }) /i\sqrt{2}$, $x_{in}=( F_{c}+F_{c}^{\dag }) /\sqrt{2}$ and $y_{in}=( F_{c}-F_{c}^{\dag }) /i\sqrt{2}$. Owing to
the Gaussian nature of the quantum noise terms $\xi $, $a_{in}$ and $F_{c}$, and to the linearization of the dynamics, the steady state of the
quantum fluctuations of the system is a CV tripartite Gaussian state, which is completely determined by the $6\times 6$ correlation matrix (CM)
$V_{ij}=\langle u_{i}( \infty ) u_{j}( \infty ) +u_{j}( \infty ) u_{i}( \infty ) \rangle /2 $. The Brownian noise $\xi(t)$ is not
delta-correlated and therefore does not describe a Markovian process~\cite{GIOV01}. However, entanglement can be achieved only with a large
mechanical quality factor, ${\cal Q}=\omega_m /\gamma_m \gg 1$. In this limit, $\xi(t)$ becomes delta-correlated~\cite{benguria}, $ \left
\langle \xi(t) \xi(t')+\xi(t') \xi(t)\right \rangle/2 \simeq \gamma_m \left(2\bar{n}+1\right) \delta(t-t')$, where $\bar{n}=\left(\exp\{\hbar
\omega_m/k_BT\}-1\right)^{-1}$ is the mean vibrational number. In this Markovian limit, the steady state CM can be derived from the following
equation \cite{prl07,parks}
\begin{equation}
AV+VA^{\intercal }=-D,  \label{Lyapunov}
\end{equation}%
where $D=Diag\left[ 0,\gamma _{m}\left( 2\bar{n}+1\right) ,\kappa ,\kappa ,\gamma _{a},\gamma _{a}\right] $ is the diffusion matrix stemming
from the noise correlations.

We have solved Eq. (\ref{Lyapunov}) for the CM $V$ in a wide range of the parameters $G_m$, $G_{a}$, $\Delta $ and $\Delta _{a}$. We have
studied first of all the stationary entanglement of the three possible bipartite subsystems, by quantifying it in terms of the logarithmic
negativity \cite{vidal02} of bimodal Gaussian states. We will denote the logarithmic negativities for the mirror-atom, atom-field and
mirror-field bimodal partitions with $E_{ma}$, $E_{af}$ and $E_{mf}$, respectively.

\begin{figure}[t]
\includegraphics[width=0.45\textwidth]{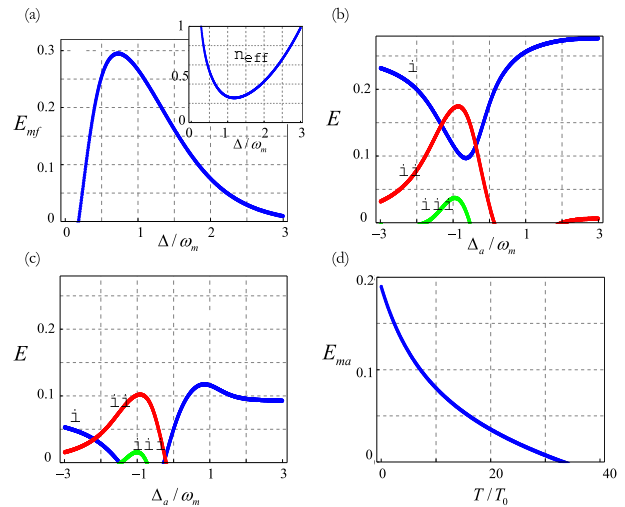}
\caption{(Color online) (a) Logarithmic negativity of the mirror-field subsystem versus the normalized cavity detuning in the absence of the
atoms. Entanglement is maximized around the optimal cooling regime (shown in the inset) namely around $\Delta \simeq \omega _{m}$ (see text for
the other parameter values). (b) Logarithmic negativity for the three bipartite entanglements as a function of the normalized atomic detuning
(see text for the value of atomic parameters). At $\Delta _{a}=-\omega _{m}$ a decrease in the mirror-field (\textit{i}, blue line) entanglement
is associated with an increase of the mirror-atoms (\textit{ii}, red line) and atoms-field (\textit{iii}, green line) entanglement. (c) Same as
in (b) but for the larger temperature $T=5T_{0}$. (d) Temperature robustness of mirror-atoms entanglement up to $40$ $T_{0}$ ($24$ K).}
\label{Plotnew}
\end{figure}

The results on the behavior of the bipartite entanglement are shown in Fig.~2. We have considered experimentally feasible parameters
\cite{gigan06,arcizet06b}, i.e., an oscillator with $\omega _{m}/2\pi =10^{7}$ Hz, $\mathcal{Q}=10^{5}$ and $m=10$ ng coupled to a cavity driven
by a laser of power $P=35$ mW at $\lambda _{l}=1064$ nm (corresponding to $G_m/2\pi =8\times 10^{6}$ Hz), with length $L=1$ mm and finesse $
\mathcal{F}=3\times 10^{4}$. The properties of the chosen working point of the cavity system are shown in Fig.~2a, showing the mirror-cavity
mode logarithmic negativity and, in the inset, the effective mean excitation number of the mechanical oscillator, $n_{eff}$, \emph{in the
absence of the atoms}, versus the normalized cavity detuning. The inset shows that we are close to ground state cavity cooling of the mirror
vibrational mode because $n_{eff}$ is decreased from the initial value $ \bar{n}=1250$ (corresponding to a reservoir temperature $T_0=0.6$ K) to
$n_{eff}\simeq 0.2$ when $\Delta=\omega_m$, i.e., the cavity is resonant with the anti-Stokes sideband of the laser. This cooling regime allows
to reach simultaneously a significant optomechanical entanglement. This can be understood in view of the results of \cite{prltelep,pirandola03},
where the entanglement between a vibrating mirror and the scattered optical sidebands is analyzed; when the mirror effective temperature is low
enough one can have strong mirror-Stokes sideband entanglement. This latter entanglement is then exploited when the atomic ensemble is placed
within the cavity. In Fig.~\ref{Plotnew}(b)-(c), the logarithmic negativity of the three bipartite cases is plotted versus the normalized atomic
detuning when $\gamma _{a}/2\pi =5\times 10^{6}$ Hz, and $G_{a}/2\pi =6\times 10^{6}$ Hz. It is evident that one has a sort of entanglement
sharing: due to the presence of the atoms, the initial cavity-mirror entanglement is partially redistributed to the atom-mirror and atom-cavity
subsystems and this effect is predominant when the atoms are resonant with the Stokes sideband ($\Delta_a=-\omega_m$). It is remarkable that, in
the chosen parameter regime, the largest stationary entanglement is the one between atoms and mirror which are only indirectly coupled.
Moreover, the nonzero atom-cavity entanglement appears only thanks to the effect of the mirror dynamics because in the bosonic approximation we
are considering and with a fixed mirror, there would be no direct atom-cavity entanglement. We also notice that atom-mirror entanglement is
instead not present at $\Delta _{a}=\omega _{m}$. This is due to the fact that the cavity-mirror entanglement is mostly carried by the Stokes
sideband and that, when $\Delta _{a}=\omega _{m}$, mirror cavity-cooling is disturbed by the Antistokes photons being recycled in the cavity by
the absorbing atoms.

Fig.~\ref{Plotnew}(c) shows the same plot but at a higher temperature, $T=5T_0=3$ K, showing that the three bipartite entanglements are quite
robust with respect to thermal noise. This is studied in more detail in Fig.~\ref{Plotnew}(d), where the atom-mirror entanglement at
$\Delta_a=-\omega_m$ is plotted versus the reservoir temperature: such an entanglement vanishes only around $20$ K.

The simultaneous presence of all the three possible instances of bipartite entanglement witnesses the strong correlation between the atoms, the
intracavity field, and the mechanical resonator at the steady state. This is also confirmed by the fact that such a state is a fully inseparable
tripartite CV entangled state in the parameter regime of Fig.~2, for a wide range of atomic detuning ($-3\omega_m < \Delta_a < 3 \omega_m$) and
up to temperatures of about $30$ K. This has been checked by applying the results of Ref.~\cite{Giedke02}, which provide a necessary and
sufficient criterion for the determination of the entanglement class of a tripartite CV Gaussian state.

We notice that the chosen parameters correspond to a small cavity mode volume ($V \simeq 10^{-12}$ m$^3$), implying that for a dipole
transition, $g$ is not small. Therefore the assumed weak coupling condition $g^2 \ll \Delta_a^2 +\gamma_a^2$ can be satisfied only if $g$
represents a much smaller, \emph{time averaged}, coupling constant. This holds for example for an atomic vapor cell much larger than the cavity
mode: if the (hot) atoms move in a cylindrical cell with axis orthogonal to the cavity axis, with diameter $\sim 0.5$ mm and height $\sim 1$ cm,
they will roughly spend only one thousandth of their time within the cavity mode region. This yields an effective $g \sim 10^4$ Hz, so that the
assumptions made here hold, and the chosen value $G_{a}/2\pi =6\times 10^{6}$ Hz can be obtained with $N_a \sim 10^7$. An alternative solution
could be choosing a cold atomic ensemble and a dipole-forbidden transition.

The entanglement properties of the steady state of the tripartite system can be verified by experimentally measuring the corresponding CM. This
can be done by combining existing experimental techniques. The cavity field quadratures can be measured directly by homodyning the cavity
output, while the mechanical position and momentum can be measured with the setup proposed in \cite{prl07}, in which by adjusting the detuning
and bandwidth of an additional adjacent cavity, both position and momentum of the mirror can be measured by homodyning the output of this second
cavity. Finally, the atomic polarization quadratures $x$ and $y$ (proportional to $S_x$ and $S_y$) can be measured by adopting the same scheme
of Ref.~\cite{sherson}, i.e., by making a Stokes parameter measurement of a laser beam, shined transversal to the cavity and to the cell and
off-resonantly tuned to another atomic transition.

In conclusion we have proposed a scheme for the realization of a hybrid quantum correlated tripartite system formed by a cavity mode, an atomic
ensemble inside it, and a vibrational mode of one cavity mirror. We have shown that, in an experimentally accessible parameter regime, the
steady state of the system shows both tripartite and bipartite CV entanglement. The realization of such a scheme will open new perspectives for
the realization of quantum interfaces and memories for CV quantum information processing and also for quantum-limited displacement measurements.

This work was supported by the European Commission (programs QAP and SCALA), and by the Italian Ministry for University and Research (PRIN-2005
2005024254).


\end{document}